# MINIMUM DISSATISFACTION PERSONNEL SCHEDULING

**Mugurel Ionut Andreica**
Politehnica University of Bucharest
mugurel.andreica@cs.pub.ro

**Romulus Andreica and Angela Andreica**
Commercial Academy Satu Mare
academiacomerciala@yahoo.com

**Abstract:** In this paper we consider two problems regarding the scheduling of available personnel in order to perform a given quantity of work, which can be arbitrarily decomposed into a sequence of activities. We are interested in schedules which minimize the overall dissatisfaction, where each employee's dissatisfaction is modeled as a time-dependent linear function. For the two situations considered we provide a detailed mathematical analysis, as well as efficient algorithms for determining optimal schedules.

## 1. Introduction

Personnel scheduling problems are very important in many activity domains and efficient scheduling techniques are being sought increasingly often. Common scheduling objectives are: maximizing productivity, minimizing losses, maximizing profit and others. In this paper we consider two problems regarding the scheduling of employees, in order to perform a given quantity of work which can be arbitrarily decomposed into a sequence of activities. The objective is to determine a time schedule of the activities and an assignment of employees to the activities which minimizes the overall dissatisfaction, subject to several types of constraints. The dissatisfaction of each employee is modeled as a time-dependent linear function. We present a detailed mathematical analysis of the problems, as well as efficient algorithms for determining optimal schedules. Our results are significant both from a theoretical and a practical point of view.

The rest of this paper is structured as follows. In Sections 2 and 3 we present the two personnel scheduling problems we mentioned, together with complete analysis and algorithms. In Section 4 we discuss related work and in Section 5 we conclude.

## 2. Minimum Dissatisfaction Scheduling with Personnel Ordering Restrictions

An economic agent has N employees, numbered with natural numbers from 1 to N, which have to perform some quantity of work. The overall quantity of work can be divided into any number of activities, which can be performed sequentially. Each activity i can be realized in negligible time (zero time), thus its only parameter of interest will be the moment when the activity is scheduled $ta_i$. Let's consider that the quantity of work has been divided into k activities, scheduled at times $0 \leq ta_1 < ta_2 < \ldots < ta_k$. The dissatisfaction of each employee j is a time-dependent linear function, $ds(j,t)=w_j \cdot |t-te_j|$, where $w_j$ is the "weight" of employee j and $te_j$ is the time moment when the employee j is the most willing (satisfied) to perform an activity (the *optimal employee time*). By |X| we denote the absolute value of X. Assuming that employee j has been assigned to activity a(j), the overall dissatisfaction will be

$$D = \sum_{j=1}^{N} ds(j, ta_{a(j)})$$

We are interested in minimizing the overall dissatisfaction. In order to do this, we can choose the total number of activities, the time moments when these activities are scheduled and the assignment of employees to activities. The only constraint we need to consider is that, if u and v are two employees such that $1 \leq u < v \leq N$, we must have $a(u) \leq a(v)$. This means that employee u cannot be assigned to an activity taking place after the activity to which employee v is assigned. Thus, the ordering of the employees must correspond to the chronological ordering of the activities to which they are assigned: $a(1) \leq a(2) \leq \ldots \leq a(N)$ ($ta_{a(1)} \leq ta_{a(2)} \leq \ldots \leq ta_{a(N)}$). We will first provide a dynamic programming algorithm for computing an optimal schedule when the maximum value of the time moments $te_j$ ($T_{max}$) is not too large and all the time moments are integer. We will compute a table $D_{min}[i,t]$=the minimum overall dissatisfaction of the employees i, i+1, ..., N, if they are assigned to activities scheduled at time moments $t' \geq t$ :

$$D_{min}[N+1, t]=0.$$

$$D_{min}[i,t] = \min_{1 \leq i \leq N} \begin{cases} D_{min}[i, t+1] \\ ds(i,t) + D_{min}[i+1, t] \end{cases}$$

In the first case, no activity is scheduled at time t, thus we will use the value $D_{min}[i,t+1]$. In the second case, we schedule an activity at time t and assign the employee i to it. The employees i+1, ..., N can be assigned to the same activity or to a subsequent one, thus we will consider the value $D_{min}[i+1,t]+ds(i,t)$. The value of t is between 0 and $T_{max}$ and the time complexity of the algorithm is $O(N \cdot T_{max})$. The minimum overall dissatisfaction is found at $D_{min}[1,0]$. The schedule can be easily computed from the table $D_{min}$, by tracing back the way the values of the table were computed.

We will now present a greedy algorithm which does not require the time moments $te_j$ to be "small". At first, we will consider that we have only one activity, scheduled at time 0 and all the employees are assigned to this activity.

Then, we will iteratively improve this solution by adding extra activities or by delaying the existing ones and reassigning the employees to these activities. We will maintain an array *tasgn*, where *tasgn[i]* is the moment when the activity to which employee is assigned is scheduled (initially, tasgn[i]=0, $1 \leq i \leq N$). If at some step of the algorithm, an employee i is assigned to an activity scheduled at time t, it will be possible to reassign this employee to an activity scheduled at a time moment t'>t (but not at a time moment t'<t). We will maintain an array *dinc*, where *dinc[j]*=the value by which the dissatisfaction of employee j increases if the employee is reassigned to an activity starting at time (tasgn[i]+1): dinc[j]=$w_j$, if $te_j \leq$tasgn[j], or $-w_j$, if $te_j>$tasgn[j]. We will compute an array *incsum*, where

$$incsum[j] = \sum_{p=j}^{N} dinc[p]$$

Thus, incsum[j] is the sum of the increases of dissatisfaction of the employees j, j+1, …, N. We will select the minimum value of the array incsum: let this value be incsum[p]. If incsum[p]$\geq$0, then the algorithm will stop. If incsum[p]<0, then, by assigning all the employees p, p+1, …, N to a an activity scheduled later than their current activity, the overall dissatisfaction will decrease. We will find the largest negative value $T_{shift}$ from the set {tasgn[q]-te$_q$ | p$\leq$q$\leq$N }, i.e. that value which is closest to 0. Then, we will increase all the values tasgn[q] (p$\leq$q$\leq$N) by |$T_{shift}$|. After doing this, we will recompute the arrays dinc and incsum and perform another iteration. The pseudocode of the algorithm is given below:

**GreedyPersonnelScheduling:**
**Step 1.** *for i=1 to N do tasgn[i]=0*
**Step 2.** *compute the arrays dinc and incsum*
**Step 3.** *choose the minimum value (incsum[p]) in the array incsum*
**Step 4. if** (*incsum[p]$\geq$0*) **then goto Step 8**.
**Step 5.** *find the minimum value $T_{shift}$=min{tasgn[q]-te$_q$ | (p$\leq$q$\leq$N) and (tasgn[q]-teq < 0) }*
**Step 6. for** *i=p to N do tasgn[p]=tasgn[p]+|$T_{shift}$|*
**Step 7. go to Step 2.**
**Step 8.** *compute* $D = \sum_{i=1}^{N} ds(i, tasgn[i])$

**Step 9. return** *D*

The algorithm above can be easily implemented in time O($N^2$). We will now explain how its time complexity can be reduced to O(N·log(N)). For this, we will use the segment tree data structure and, in particular, the segment tree framework introduced in [1]. We will maintain three segment trees, as showed in the pseudocode below:

**GreedyPersonnelScheduling-O(N·logN):**
**for** *i=1* **to** *N* **do**
 *tasgn[i]=0*
 **if** (*te$_i\leq$tasgn[i]*) **then** *dinc[i]=w$_i$*
 **else** *dinc[i]=-w$_i$*
*incsum[N]=dinc[N]*
**for** *i=N-1* **downto** *1* **do**
 *incsum[i]=dinc[i]+incsum[i-1]*
// compute two auxiliary arrays: incsum_aux and tasgn_aux
**for** *i=1* **to** *N* **do**
 *incsum_aux[i]=(incsum[i], i)*
 **if** (*dinc[i]$\geq$0*) **then** *tasgn_aux[i]=(-∞, i)*
 **else** *tasgn_aux[i]=(tasgn[i]-te$_i$, i)*
*st_iaux = the segment tree for the array incsum_aux, with update function uFunc=plus_poz and query function qFunc=min_poz*
*st_taux = the segment tree for the array tasgn_aux, with update function uFunc=plus_poz and query function qFunc=max_poz*
*st_tasgn = the segment tree for the array tasgn, with update function uFunc=+ and query function qFunc=min*
**while** (*true*) **do**
 (*vmin, pozmin*)=**STrangeQuery**(*st_iaux.root, 1, N*)
 **if** (*vmin$\geq$0*) **then break**
 ($T_{shift}$, *pozshift*)=**STrangeQuery**(*st_taux.root, pozmin, N*)
 **STrangeUpdate**(*st_taux.root, (-Tshift, 0), pozmin, N*)
 **STrangeUpdate**(*st_tasgn.root, -Tshift, pozmin, N*)
 **STrangeUpdate**(*st_taux.root, (-∞, 0), pozshift, pozshift*)
 **STrangeUpdate**(*st_iaux.root, (2·w$_{pozshift}$, 0), 1, pozshift*)
**for** *i=1* **to** *N* **do**
 *tasgn[i]=***STrangeQuery**(*st_tasgn.root, i, i*)

The functions *plus_poz*, *min_poz* and *max_poz* are defined below:

**plus_poz((v$_x$, poz$_x$), (v$_y$, poz$_y$)):**
**return** ((*v$_x$+v$_y$*), **max**(*poz$_x$, poz$_y$*))

**min_poz((v$_x$, poz$_x$), (v$_y$, poz$_y$)):**
**if** (*v$_x$<v$_y$*) **then return** (*v$_x$, poz$_x$*)
**else return** (*v$_y$, poz$_y$*)

**max_poz((v$_x$, poz$_x$), (v$_y$, poz$_y$)):**
**if** (*v$_x$>v$_y$*) **then return** (*v$_x$, poz$_x$*)
**else return** (*v$_y$, poz$_y$*)

The functions *STrangeUpdate* and *STrangeQuery* are part of the segment tree framework defined in [1]. The algorithm works as follows: we find the minimum value of the array incsum (vmin) and the position of the minimum value (pozmin) using the segment tree *st_iaux*. If vmin$\geq$0, then the execution ends. Afterwards, we find the largest non-positive value $T_{shift}$ in the array tasgn_aux, together with its position *pozshift*, using the segment tree *st_taux*. Then, we increase by |$T_{shift}$| all the positions in the arrays *tasgn_aux* and *tasgn*, between *pozmin* and *N*. We set the value of *tasgn_aux[pozshift]* to (-∞, *pozshift*), in order to ignore this position from now on. After all these operations, the value *dinc[pozshift]* changes from -w$_{pozshift}$ to +w$_{pozshift}$. Thus, all the values incsum[p] (1$\leq$p$\leq$pozshift) increase by 2·w$_{pozshift}$. All the operations are performed in O(log(N)) time per iteration and the algorithm performs O(N) iterations, arriving at a time complexity of O(N·log(N)).

## 3. Minimum Dissatisfaction Scheduling with Increasing Optimal Employee Times

This situation is similar to the previous one, except that the optimal employee times te$_1$, te$_2$, …, te$_N$ are sorted in increasing order, i.e. te$_1\leq$te$_2\leq$…$\leq$te$_N$. There are also other restrictions: the number of activities is fixed to a given value k and they can be scheduled only at time moments equal to optimal employee time moments. Furthermore, any two activities must be scheduled at different time moments. We will enhance the model by considering the dissatisfaction of the employer in the following way: if an activity is scheduled at time moment te$_j$, then the employer's dissatisfaction will be de$_j\geq$0. The objective is to minimize the overall dissatisfaction (the dissatisfaction of the employees plus the dissatisfaction of the employer). We will solve this problem by dynamic programming. We will compute two sets of values: D$_{min}$[i,j,0] and D$_{min}$[i,j,1]:

- D$_{min}$[i,j,0]=the minimum overall dissatisfaction if the i$^{th}$ activity is scheduled at time te$_j$ (and all the

employees 1,2,...,j are assigned to one of the activities 1,2,...,i)
- $D_{min}[i,j,1]$=the minimum overall dissatisfaction if the $i^{th}$ activity is scheduled at a time moment $t \leq te_j$ (and all the employees 1,2,...,j are assigned to one of the activities 1,2,...,i)

We have the following recurrence equations:

$$D_{\min}[i, j, 0] = \min_{0 \leq j' < j} \{D_{\min}[i-1, j', 1] + de_j + \sum_{p=j'+1}^{j} w_p \cdot (te_j - te_p)\}$$

$$D_{\min}[i, j, 1] = \min_{1 \leq j' \leq j} \{D_{\min}[i, j', 0] + \sum_{p=j'+1}^{j} w_p \cdot (te_p - te_{j'})\}$$

The initial values are: $D_{min}[0,0,0]=D_{min}[0,0,1]=0$ and $D_{min}[0,j,0]=D_{min}[0,j,1]=+\infty$ (for $j>0$). The minimum overall dissatisfaction is equal to $D_{min}[k,N,1]$ and the activity schedule and assignment of employees to activities can be determined by tracing back the way the $D_{min}[i,j,p]$ values were computed. A naive algorithm implements the equations directly and has time complexity $O(N^2 \cdot k)$, considering that we can evaluate in $O(1)$ time the sums with the p argument. We can achieve this by computing the arrays *wsum*, *wright* and *wleft*:

- *wsum[i]* = the sum of the weights of the employees 1,2,...,i; wsum[0]=0 and wsum[i] = wsum[i-1]+w[i]
- *wright[i]* = the total dissatisfaction of the employees i, i+1, ..., N, if they are assigned to an activity scheduled at time $te_N$; wright[N+1]=0 ; wright[i] = wright[i+1]+ $w_i \cdot (te_N - te_i)$
- *wleft[i]* = the total dissatisfaction of the employees 1, 2,...,i, if they are assigned to an activity scheduled at time $te_1$; wleft[0]=0 ; wleft[i] = wleft[i-1] + $w_i \cdot (te_i - te_1)$

With these arrays, we can write $\sum_{p=j'+1}^{j} w_p \cdot (te_j - te_p)$ as (wright[j'+1] – wright[j+1]– (wsum[j]–wsum[j'])·($te_N$-$te_j$)). Similarly, $\sum_{p=j'+1}^{j} w_p \cdot (te_p - te_{j'})$ is equal to (wleft[j]-wleft[j']-(wsum[j]-wsum[j'])·($te_{j'}$-$te_1$)).

We can improve the algorithm to $O(N \cdot K)$, by introducing the following concepts: for each activity i ($1 \leq i \leq k$) and each employee j ($1 \leq j \leq N$), we will define two functions: $f_{i,j}$ and $g_{i,j}$, which will be used in order to compute the values $D_{min}[i,j,0]$ and $D_{min}[i,j,1]$. The functions $f_{i,j}$ are defined on the interval $[te_j, te_N]$. $f_{i,j}(te_p)$ represents the minimum dissatisfaction of the employees 1,2,...,p if i activities were scheduled, the $i^{th}$ activity is scheduled at time $te_p$ and the employees j, j+1, .., p are assigned to activity i. With this definition, $D_{min}[i,j,0]$ is the minimum value $f_{i,j'}(te_j)$ ($0 \leq j' < j$), plus $de_j$. The important issue now becomes to find the minimum value of these functions, without evaluating every function at the time moment $te_j$. (which would get us back to an $O(N^2 \cdot k)$ algorithm).
The equation of a function $f_{i,j}(te_p)$ is: $D_{min}[i-1, j-1, 1]$ + $\sum_{q=j}^{p} w_q \cdot (te_p - te_q)$ . The difference between two "consecutive" values of a function $f_{i,j}$ is:

$df_{i,j}(te_{p+1}) = f_{i,j}(te_{p+1}) - f_{i,j}(te_p) = (w[j]+w[j+1] + ... + w[p]) \cdot (te_{p+1} - te_p)$. We notice that $df_{i,j}(te_{p+1}) < df_{i,j'}(te_{p+1})$, when $j' < j$ (the functions which "started" more recently grow slower than those which have "started" for a longer time). This is because the sum $(w[j]+...+w[p])$ is larger when j is smaller. From this observation we conclude the following:

- if the value of a function $f_{i,j}(te_{p+1})$ is larger than the value of a function $f_{i,j'}(te_{p+1})$, with $j'>j$, then the function $f_{i,j}$ will never have the minimum value (among all the functions) at any of the subsequent steps.
- if the value of a function $f_{i,j}(te_{p+1})$ is smaller than the value of a function $f_{i,j'}(te_{p+1})$, with $j'>j$, then the function $f_{i,j}$ will be "surpassed" by the function $f_{i,j'}$ at a time moment $t_{surpass,i,j,j'}$; $f_{i,j'}$ will not have the minimum value among all the functions $f_{i,j}$ ($j \leq j'$) before a time moment equal to the maximum value of the set $\{t_{surpass,i,j,j'}|j<j'\}$.

We can use a double-ended queue (deque), in order to store all the functions which "started" up to a time step p (time moment $te_p$). Within the deque, the functions are sorted according to their value at step p, as well as after the time moment when their value will be the minimum one among all the functions which "started" before them ($t_{early,i,j}$ for a function $f_{i,j}$). At every step p, a new function $f_{i,p}$ is inserted into the deque. This function will remove from the end of the deque all the functions having a value which is larger than $f_{i,p}$ at time moment $te_p$, as well as those functions j for which $t_{surpass,i,j,p}$ is smaller than $t_{early,i,j}$ (because these function will be surpassed by the function $f_{i,p}$ before getting the chance to have the minimum value among all the other functions; thus, their values wil never be globally minimum). Moreover, at every step p, we iteratively remove the first function from the front of the deque if the second function $f_{i,j}$ has $t_{early,i,j} < te_p$.

In order to compute the time moment $t_{surpass,i,j,j'}$ when a function $f_{i,j'}$ surpasses a function $f_{i,j}$ ($j < j'$), we must compute the following values. Let's assume that we are at time step p=j'. We will compute dC= $f_{i,j}(te_{j'}) - f_{i,j'}(te_{j'})$. We notice that in between two steps j'-1 and j', the functions $f_{i,j}$ behave like half-lines, with a slope equal to $dP_j = (w_j + w_{j+1} + ... + w_{j'-1})$. The slope of function $f_{i,j'}$ is $dP_{j'}=0$. At every time moment after $te_{j'}$, the difference between the slopes of the two functions remains constants and equal to $dP_j - dP_{j'}$. This is easily noticeable, because the slopes of the two functions will increase with the same amount at every step q>j'. Thus, the time moment when the function $f_{i,j'}$ surpasses the function $f_{i,j}$ is $t_{surpass,i,j,j'} = te_{j'} + dC/(dP_j - dP_{j'})$.

In order to compute the values $D_{min}[i,j,1]$, we will proceed in a similar manner. We will define some functions $g_{i,j}$:[wsum[j], wsum[N]], whose values $g_{i,j}$(wsum[p]) represent the minimum dissatisfaction of the employees 1,2,...,p, if the $i^{th}$ activity is scheduled at time $te_j$ and the employees j, j+1, ...,p are assigned to activity i. These functions are defined on the partial sums of the weights of the employees, in order to be able to use a similar reasoning. Every function $g_{i,j}$ will be a half-line (with constant slope) in between two "consecutive" points wsum[j'-1] and wsum[j']. The slope of a function $g_{i,j}$ will be, according to this definition, equal to $te_{j'} - te_j$. The pseudocode of the algorithm is given below:

**DPPersonnelScheduling:**
*compute the arrays wsum, wright and wleft*
*initialize $D_{min}[0,j,0]$ and $D_{min}[0,j,1]$ ($0 \leq j \leq N$)*
**for** *i=1* **to** *k* **do**
  $dq_0$=empty; $dq_1$=empty
  **for** *j=0* **to** *k-1* **do**
    $D_{min}[i,j,0]=D_{min}[i,j,1]=+\infty$
  **for** *j=k* **to** *N* **do**
    // clean up the front of dq0
    **while** ($dq_0$.size()>1) **and** ($dq0$.getSecond().$t_{early}$<$te_j$) **do**
      $dq_0$.removeFirst()
    // compute Dmin[i,j,0]
    $t_{early,i,j}=te_j$
    **while** ($dq_0$.size()≥1) **do**
      $e=dq_0$.getLast()
      $x=e.v+wright[e.j]-wright[j+1]-$
                    $(wsum[j]-wsum[e.j-1])\cdot(te_N-te_j)$
      $dC=D_{min}[i-1,j-1,1]-x$
      $dP=wsum[j-1]-wsum[e.j-1]$
      **if** ($dC\leq 0$) **then** $t_{surpass,i,e,j,j}=-\infty$
      **else** $t_{surpass,i,e,j,j}=te_j+(dC/dP)$
      **if** ($t_{surpass,i,e,j,j} \leq e.t_{early}$) **then**
        $dq_0$.removeLast()
      **else**
        $t_{early,i,j}=t_{surpass,i,e,j,j}$
        **break**
    $dq_0$.addLast(('v'=$D_{min}[i-1,j-1,1]$, '$t_{early}$'=$t_{early,i,j}$, 'j'=j))
    $e=dq_0$.getFirst()
    $D_{min}[i,j,0]=e.v+wright[e.j]-wright[j+1]-$
                    $(wsum[j]-wsum[e.j-1])\cdot(te_N-te_j)+de_j$
    // clean up the front of dq1
    **while** ($dq_1$.size()>1) **and** ($dq_1$.getSecond().$w_e$<$wsum[j]$) **do**
      $dq_1$.removeFirst()
    // compute Dmin[i,j,1]
    $w_{early,i,j}=wsum[i]$
    **while** ($dq_1$.size()≥1) **do**
      $e=dq_1$.getLast()
      $x=e.v+wleft[j]-wleft[e.j]-(wsum[j]-wsum[e.j])\cdot(te_{e.j}-te_1)$
      $dC=D_{min}[i,j,0]-x$
      $dP=te_j-te_{e.j}$
      **if** ($dC\leq 0$) **then** $w_{surpass,i,e,j,j}=-\infty$
      **else** $w_{surpass,i,e,j,j}=wsum[j]+(dC/dP)$
      **if** ($w_{surpass,i,e,j,j} \leq e.w_e$) **then**
        $dq_1$.removeLast()
      **else**
        $w_{early,i,j}=w_{surpass,i,e,j,j}$
        **break**
    $dq_1$.addLast(('v'=$D_{min}[i,j,0]$, '$w_e$'=$w_{early,i,j}$, 'j'=j))
    $e=dq_1$.getFirst()
    $D_{min}[i,j,1]=e.v+wleft[j]-wleft[e.j]-(wsum[j]-wsum[e.j])\cdot(te_{e.j}-te_1)$

The algorithm has $O(N \cdot k)$ amortized complexity. The key element of the algorithm is the deque data structure. At every step (i,j), many operations can be performed on the deque, but only O(N) operations are performed on the deque for a given value of i (and all the values of j).

## 4. Related Work

Personnel scheduling is an important research topic and many papers have addressed such scheduling problems, using a large variety of techniques: genetic algorithms [2], memetic algorithms [3], tabu search [4], heuristics [5], branch and price [9], integer and network programming [6]. Some techniques from other scheduling domains could also be applied, like greedy and dynamic programming algorithms [7] and efficient data structures [1]. Given a different meaning to the problem parameters, our second scheduling problem is nearly identical to the K-Median problem of a set of points on a line, which was solved in $O(N \cdot k)$ time [8].

## 5. Conclusions and Future Work

In this paper we considered two personnel scheduling problems, in which the objective consisted of minimizing the dissatisfaction of the employees, when they have to perform a sequence of activities. The dissatisfaction of each employee was modeled as a time-dependent linear function. The scheduling constraints consisted either of personnel ordering restrictions or a fixed number of activities which needed to be executed. For both problems we presented efficient algorithms for determining optimal schedules. As future work, we intend to adopt more complex dissatisfaction models and consider some multi-criteria optimization problems.